\documentclass[onecolumn,showpacs,preprintnumbers,floatfix,amsmath,amssymb,aps]{revtex4}
\usepackage{bm}
\usepackage{color}

\usepackage{graphicx,subfig,epsfig,epsf,color}
\usepackage[utf8x]{inputenc}
\usepackage{amssymb,amsmath,appendix}
\usepackage{slashed}
\usepackage{array}
\usepackage{hyperref}
\usepackage{ulem}

\begin{document}
\title{A Detailed Analysis of the Special Points on $M-R$ Solutions of Hybrid (Twin) Stars}

\author{Debashree Sen$^1$, Naosad Alam$^1$, and Gargi
Chaudhuri$^{1,2}$}

\address{$^1$Physics Group, Variable Energy Cyclotron Centre, 1/AF Bidhan Nagar, Kolkata 700064, India}
\address{$^2$Homi Bhabha National Institute, Training School Complex, Anushakti Nagar, Mumbai 400085, India}

\email{debashreesen88@gmail.com, naosadphy@gmail.com, gargi@vecc.gov.in}

\date{\today}




\begin{abstract}

Hadron-quark phase transition in neutron star cores is achieved in the present work with the help of Maxwell construction. For the purpose we employ six different and well-known hadronic models for the pure hadronic phase. The quark phase is described with the MIT Bag model in which the density dependence of the bag pressure $B(\rho)$ is invoked for different asymptotic values ($B_{as}$) of $B(\rho)$. The resulting hybrid star (HS) configurations exhibit twin star characteristics and distinct special points (SPs) on the mass-radius diagram of the HSs irrespective of the transition densities and the value of $B_{as}$. We find that for any particular value of $B_{as}$, the mass corresponding to SP ($M_{SP}$) and the maximum mass ($M_{max}$) of the HSs, obtained with different hadronic models, follow a nearly linear (fitted) relationship where the slope is independent of the value of $B_{as}$. The $M_{SP}-M_{max}$ dependence of the HSs is found to be consistent with any hadronic equation of state (EoS) chosen to obtain the hybrid EoS and thus such relations can be considered as universal relations in the context of formation of SPs. A change in the value of $B_{as}$ shifts the position of the fitted line in the $M_{SP}-M_{max}$ plane, with the linearity, however, retained. 


\end{abstract}




\maketitle



\section{Introduction}
\label{Intro}
 
 Theoretical modeling of neutron star (NS) matter (NSM) is the best way to understand the composition and interaction of matter at high density (5 - 10 times nuclear density $\rho_0$) in the absence of any conclusive knowledge from experimental perspectives. The equation of state (EoS) of the NSM is constrained to certain extent by certain astrophysical and observational results. Such constraints include those on the maximum mass of the NSs obtained from high mass pulsars like PSR J0348+0432 \cite{Ant} and PSR J0740+6620 \cite{Fonseca2021}. Recently, NICER experiment also put constraints on the radius of PSR J0740+6620 \cite{Miller2021,Riley2021}. The limit on the dimensionless tidal deformability of a 1.4 $M_{\odot}$ NS is set from the GW170817 observational data \cite{GW170817}. Other constraints on the mass-radius ($M-R$) plane of the NSs are prescribed from GW170817 data analysis \cite{GW170817} and also from recent NICER experiments for PSR J0030+0451  \cite{Riley2019,Miller2019}.
 
 With the theoretical modeling of NSM, the EoS is thus dependent on the composition and interactions considered. Theoretically, the dense environment of the NS core can support the formation of different stable exotic matter \cite{Glendenning}. It is often speculated that at such densities, deconfinement of hadronic matter may occur to form quark matter via phase transition and thereby forming hybrid stars (HSs) \cite{Glendenning,Blaschke2,Weissenborn2011,Ozel2010,Klahn, Bonanno,Drago2016,Zdunik,Wu,Schramm,Lenzi, Bhattacharya,Gomes2019,Han,Ferreira2020,Most2,Zha,Khanmohamadi, Xia2,Maslov,Dexheimer2,Montana,Christian,Contrera,Lugones2021, Bozzola,Liu2022,Tmurbagan,Steiner,Prakash,Yazdizadeh,Miyatsu,Liu, Sen,Sen2,Sen6,Sen7,Sen8}. Not only in compact star cores, the phenomenon of hadron-quark phase transition finds its application in various interesting contexts like the heavy-ion collision physics, supernova explosions and binary neutron star mergers (BNSMs) etc. Quantum Chromodynamics (QCD) calculations help us to speculate that at high temperature (as in case of the early stages of the universe) and at high density (as in case of compact stars) the formation of quark–gluon plasma (QGP). This helps to picturize the QCD phase diagram which provides the notion of hadron-quark phase transition along the chemical potential (density) and temperature axes. The high temperature - low baryon density regime of the QCD phase diagram is accessible to the heavy-ion collision experiments to certain extent and the first-principle calculations of QCD predict a smooth crossover transition from hadronic to deconfined quark phase at around T=(156.5 $\pm$ 1.5) MeV \cite{Bazavov}. The low (negligible) temperature – high density regime conditions are highly challenging to be  accessible in experiments. Such conditions prevail ideally in the core of neutron/compact stars. However, due to lack of proper understanding of matter and its interaction at such high density, the composition and the possible presence of exotic matter inside the neutron/compact star core still remain one of the most interesting and unsolved aspects of dense matter physics. Whether quark matter can be a possible candidate at such dense environment is still an open question and one of the current topics of interest. At high densities relevant to the core of compact stars and the asymptotic freedom of QCD indicate a possible transition of phase from hadronic matter to quark matter under such conditions, likely in the form of a first order phase transition \cite{Glendenning}. In other words, deconfinement of hadronic matter to quark matter is expected at such extreme conditions of density at the core of NSs thereby forming HSs. Moreover, it is also suggested that when density increases, strong first order phase transitions with large density jumps can trigger supernova explosions via propagation of shock waves. This ultimately results in the formation of proto-neutron stars inside the collapsing star’s core \cite{SN}. Also in the era of the BNSM detection, the phenomena of phase transition has gained special attention and interest. Till date the detection of the inspiralling phase of the merger specially in case of GW170817 helped to constrain the radius and tidal deformability of a 1.4 $M_{\odot}$ NS. Unfortunately, the post-merger phase could not be detected for GW170817 event. However, it has been suggested that if the post-merger phase can be detected in future then the peak frequency of the ringdown signal can disclose further information about compact star properties. The joint information from both the inspiraling and post-merger phases may also be helpful to understand the  possibility of first-order phase transition in the NS merger \cite{BNSM}.
 
 In case of HSs the hadron-quark phase transition is achieved generally with the help of Gibbs and/or Maxwell constructions depending on the value of surface tension at the hadron-quark interface. The former is based on global charge neutrality condition and the formation of mixed phase \cite{Glendenning,Bhattacharya,Sen,Sen6} while the later is characterized by density jump and the local charge neutrality condition is considered \cite{Schramm,Lenzi,Bhattacharya,Gomes2019,Han,Ferreira2020, Khanmohamadi,Sen,Sen2,Sen7}. It is well-known that if the surface tension at the boundary is too high, the mixed phase becomes unstable and then Maxwell construction is favored \cite{Maruyama}. However, actual the value of surface tension at the boundary is not known and in the present work we assume it to be high enough to invoke phase transition using Maxwell construction. For the quark phase, we consider the MIT bag model \cite{Chodos}. As the quarks acquire asymptotic degree of freedom at high densities relevant to HS cores, in the present work we consider the density dependence of the bag pressure following a Gaussian distribution form \cite{Burgio1,Burgio2}. We obtain the EoS of HSs and consequently their structural properties for different values of $B_{as}$ of the bag pressure $B(\rho)$ where, $B_{as}$ is the value of $B(\rho)$ where the quarks become asymptotic. The chosen values of $B_{as}$ are consistent with that prescribed by \cite{B_limit,Nandi} using specific models in the light of GW170817 data. For the pure hadronic phase we adopt six different relativistic mean field (RMF) models. As stated earlier, the EoS of compact star is in general dependent on its composition which is largely unknown at present from experimental perspectives. Theoretically, at high density relevant to NS cores, there may be possibility of existence of exotic forms of matter like the hyperons, delta baryons, paired and unpaired quarks and boson condensates etc. However, at present the precise conditions of temperature, density and iso-spin asymmetry are not well-known which broadly regulate the threshold for the appearance or disappearance of these exotics \cite{Blaschke2015}. Considering the hadronic composition, theoretically, the hyperons and delta baryons may appear when the neutron chemical potential matches with and surpasses the rest mass of the hyperons. Therefore several theoretical models suggest different threshold of their appearance. Moreover, the hyperon couplings in the hadronic sector are still not well-determined and is chosen on the basis of the potential depths of individual hyperon species. Among all the hyperon species, the only value of the potential depth which is reasonably known is that of the $\Lambda$ hyperon \cite{Logoteta}. Also their presence is known to soften the EoS and reduce the maximum mass of the NSs \cite{Glendenning,Weissenborn12,Sen2,Sen6,Sen3,Logoteta}. This leads to the well-known hyperon/delta puzzle. Many works have suggested various ways to solve the puzzle with mechanisms like considering the effect of repulsive hyperon-hyperon interaction via exchange of mesons \cite{Weissenborn12}, inclusion of repulsive hyperonic three-body forces \cite{Yamamoto} and invoking phase transition from hadronic to deconfined quark matter forming HSs \cite{Weissenborn2011,Klahn,Bonanno,Drago2016,Zdunik,Wu}. Now considering quarks as constituents of compact stars via phase transition, their threshold density is also inconclusive and several theoretical models have suggested different values. Few works \cite{Blaschke2015} have suggested that in case of HSs under certain circumstances, the threshold density of appearance of hyperons and quarks are often very close or overlapping. Therefore, considering all the above facts and for simplicity, we do not consider the presence of hyperons in the hadronic phase of the present work similar to \cite{Ju,Liu2022,Agrawal,Sen8}.
 
 With the obtained hybrid EoS, we compute the structural properties of the HSs in static conditions. In this context, we obtain twin star configurations and special points (SPs) on the $M-R$ diagram with all the considered hadronic models. Several works have suggested that phase transition from hadronic to unpaired or color-flavor locked quark matter often leads to the formation of a “third family” of compact stars \cite{Glendenning,Schramm,Maslov,Schertler,Blaschke2015,Alvarez-Castillo,Blaschke,Cierniak,Sharifi,Ayriyan,Jakobus,Espino, Hempel,Banik,Wang} and twin star configurations under certain circumstances of strong first order phase transitions mainly with considerable density jumps. The $M-R$ plot in such cases often show non-identical branches with two distinct maxima at two different radii. In other words twin stars are two separate points with same mass but different radii on the mass-radius diagram of the HSs. Usually one of these twins is a regular neutron (hadronic) star located on the first stable branch whereas the other is a HS lying on the second stable branch, disconnected from the first (hadronic) branch by an instable region. However, depending on its transition density, the twins with nearly identical mass can both be located in the second stable branch \cite{Lyra}. Few works have classified such twin stars broadly into four categories depending on the location of these twins in terms of mass \cite{Montana,Christian,Sharifi}. A SP ($M_{SP}$, $R_{SP}$) on the $M-R$ plot indicates the small region where all the HS solutions merge irrespective of the different transition densities for different values of bag pressure. This feature is of great interest in the context of formation of hybrid and twin stars \cite{Cierniak,Yudin,Cierniak2}. It is particularly pronounced in case of HSs with strong phase transitions that exhibit third family solutions and the twin star phenomenon \cite{Kaltenborn,Blaschke3}. There are a few works that have already obtained the SP feature of HSs using some more realistic effective quark models. This feature was first identified in \cite{Yudin} for the generic constant-speed-of-sound (CSS) quark model. Later on \cite{Cierniak,Cierniak2} with the CSS model found that the existence of SPs can be treated as universal property of HS models because their location on the mass-radius diagram is insensitive to the transition density. Hence SPs serve as a remarkable tool to interpret the recent multi-messenger observational results as signals for the possible existence of HS branches \cite{Cierniak,Cierniak2}. In \cite{Cierniak} even advanced MIT Bag models like the vector Bag (vBag) model \cite{vBag} were also adopted that yielded SPs on the mass-radius diagrams of HSs. Ref. \cite{Blaschke} adopted the color superconducting generalized non-local Nambu-Jona-Lasinio (nlNJL) model for the quark phase and obtained SPs on the mass-radius diagrams of HSs using both Maxwell construction and an interpolation procedure with a polynomial function while \cite{Ayriyan} obtained SPs with both Maxwell construction and interpolation procedure with a polynomial function in terms of a mixed phase parameter. Ref. \cite{Ivanytskyi} adopted the relativistic density-functional approach \cite{Kaltenborn} to model the color superconducting quark matter and studied the properties of HSs. They obtained SPs for the variation of the pairing or gap parameter and the effective quark mass. Ref. \cite{Wang} studied smooth phase transition with Gibbs construction and obtained SPs on the mass-radius diagram for different values of the bag constant. In the present work with the considered number of hadronic models and for different values of $B_{as}$ for density-dependent bag model, we have achieved phase transition with Maxwell construction and analyzed the properties of HSs. The location of the SPs obtained in this work is compared with the various recent constraints on the mass-radius relationship of compact stars. We have also constrained the value of the maximum gravitational mass $M_{max}$ of the HSs with respect to $M_{SP}$. As we obtain a linear $M_{max}$-$M_{SP}$ involving the six hadronic models, these relations can be treated as universal relations in the context of existence of SPs.
 
 The paper is organized as follows. In the next section \ref{Formalism}, we address the six hadronic models adopted (Section \ref{Hadronic_model}). In Section \ref{Quark phase}, the main features of the density dependent bag model for the pure quark phase are highlighted along with the mechanism of phase transition with Maxwell construction. We then present our results and relevant discussions in section \ref{Results}. We summarize and conclude in the final section \ref{Conclusion} of the paper.
  

\section{Formalism}
\label{Formalism}

\subsection{Pure Hadronic Phase}
\label{Hadronic_model}

 For the pure hadronic phase, we employ six different RMF models viz. TM1 \cite{tm1} BSR2, BSR6 \cite{bsr}, GM1 \cite{gm1}, NL3$\omega\rho$4 \cite{Pais} and NL3 \cite{nl3}. The saturation properties of these models differ from each other and are in reasonable agreement with the different experimental and empirical data. In table \ref{tab:1} we list the saturation properties like the saturation density ($\rho_0$), binding energy per particle ($e_0$), nuclear incompressibility ($K_0$), symmetry energy coefficient ($J_0$) and the slope parameter ($L_0$) of the chosen hadronic models.

\begin{table}[!ht]
\caption{The nuclear matter properties at saturation density $\rho_{_{0}}$ for different hadronic models.}
\setlength{\tabcolsep}{12.0pt}
\begin{tabular}{cccccc}
 \hline
 \hline
  Model & $\rho_{0}$ & $e_{0}$ & $K_{0}$ & $J_{0}$ & $L_{0}$ \\
  & $({\rm fm}^{-3})$ & (MeV) & (MeV) & (MeV) & (MeV) \\ \hline
TM1 & 0.145  & $-$16.26 & 281.2 & 36.9 & 110.8  \\
BSR2 & 0.149 & $-$16.03 & 240.0 & 31.4 & 62.2   \\
BSR6 & 0.149 & $-$16.13 & 235.9 & 35.4 & 85.6   \\
GM1  & 0.153 & $-$16.30 & 300.1 & 32.5 & 93.9   \\
NL3$\omega\rho$4 & 0.148 & $-$16.25 & 271.6 & 33.1 & 68.2 \\
NL3 & 0.148 & $-$16.25 & 271.6 & 37.4 & 118.5 \\
\hline
\hline
\end{tabular}
\label{tab:1}
\end{table}

 The symmetry energy coefficient ($J_0$) and the slope parameter ($L_0$) of the chosen hadronic models are quite consistent with the recent findings of \cite{Reed} obtained from the correlation between them and the neutron skin thickness of $^{208}\rm{Pb}$ ($R^{208}_{skin}$) as measured by the PREX-II experiment. The binding energy per particle ($e_0$) and the saturation density ($\rho_0$) of the different hadronic models are also consistent with the phenomenological analysis of \cite{Dutra}. However, the saturation density of TM1 is slightly less than that prescribed in \cite{Dutra}. In absence of any direct experimental determination of energy per particle and saturation density, they are extracted from certain experimental data analysis \cite{Margueron,expt,expt2}. Ref. \cite{Margueron} has presented a helpful list of references in this regard. The nuclear incompressibility ($K_0$) of certain hadronic models like TM1, GM1, NL3$\omega\rho$4 and NL3 are slightly larger than the prescribed experimental finding of \cite{K} but consistent with \cite{Stone}. The value of $K_0$ in such works have a wide range of uncertainties pertaining to it and the value of $K_0$ is still not an experimentally well-measured quantity. All the chosen hadronic models in this work are quite well-known and have been extensively adopted in literature, even in recent works, to determine the properties of neutron/hybrid stars.

 As mentioned in section \ref{Intro} we do not include the hyperons and the delta baryons in the hadronic sector because these heavier baryons soften the EoS which results in low mass of the HSs \cite{Glendenning,Weissenborn12,Sen2,Sen6,Sen3}. Also their couplings are not experimentally well-known except for that of the $\Lambda$ hyperon \cite{Logoteta}. Therefore, similar to works like \cite{Ju,Liu2022,Agrawal,Sen8} in the context of phase transition, we consider $\beta$ equilibrated matter consisting of the nucleons, electrons and muons as the composition of the hadronic phase described with the six different RMF models.

\subsection{Pure Quark Phase and Hadron-Quark Phase Transition}
\label{Quark phase}

 We adopt the MIT Bag model \cite{Chodos} with u, d and s quarks along with the electrons to describe the pure quark phase. The mass of the u and d quarks are quite small compared to that of the s quark ($m_s \approx$ 95~MeV). The model is based on the hypothesis that the unpaired quarks are constrained within a hypothetical region known as the `Bag', characterized by specific bag pressure $B$ that determines the strength of quark interaction. This bag pressure signifies the difference in energy density between the perturbative vacuum and the true vacuum \cite{Burgio1,Burgio2}. The value of $B$ is still inconclusive and it is often taken as free parameter that plays an important role in determining the properties of the HSs. In the light of the constraints from GW170817 observation, \cite{B_limit,Nandi} have put limits on the value of $B$ for HSs considering a few well-known hadronic models for the hadronic phase.

 It is well-known that the quarks at high densities, relevant to NS/HS cores, enjoy asymptotic freedom \cite{Burgio1,Burgio2}. This fact justifies that the bag pressure to be density dependent rather than being a constant. Therefore in the present work we consider the density dependence of the bag pressure $B(\rho)$ following a Gaussian distribution form \cite{Burgio1,Burgio2} given as

\begin{eqnarray}
B(\rho) = B_{as} + (B_0 - B_{as})~ \rm{exp}~ [-\beta(\rho/\rho_0)^2]
\label{B}
\end{eqnarray}

where, $B_0$ and $B_{as}$ are the values attained by $B(\rho)$ at $\rho=0$ and asymptotic densities, respectively. $\beta$ controls the decrease of $B(\rho)$ with the increase of density. Such a distribution form, regulating the density dependence of the bag pressure, involves the asymptotic behavior of the quarks at high densities relevant to HS cores. It is already shown that this can significantly affect the structural properties of HSs \cite{Burgio1,Burgio2,Sen7}. In the present work, we choose $B_0=400$ MeV fm$^{-3}$ and $\beta=0.17$ following \cite{Burgio1,Burgio2}. As hadron-quark phase transition is expected at high densities, therefore the precise value of $B_0$ is not important in the present context. However, $B_{as}$ is of greater significance and relevance in case of HSs. Thus to obtain the hybrid EoS, we vary $B_{as}$ consistent with the limits proposed by \cite{B_limit,Nandi}. In the present work, we consider the simplistic form of the Bag model without involving the strong repulsive interactions between the quarks. First order correction due to strong interaction \cite{Fraga} and perturbative effects \cite{Weissenborn2011,Benhar,Nakazato,Uechi,Bombaci17} may also be considered. However, \cite{Glendenning,Bhattacharya,Tmurbagan,Steiner,Prakash, Yazdizadeh,Miyatsu,Liu} noted that the effects of perturbative corrections can also be realized by varying the bag pressure. 

 In the MIT Bag model, the energy density and pressure of the quarks can be expressed as \cite{Glendenning}

\begin{eqnarray}
\varepsilon_{Q} = B(\rho)+ \sum_f \frac{3}{4\pi^2} \Biggl[\mu_fk_f\Biggl(\mu_f^2-\frac{1}{2}m_f^2\Biggr) - \frac{1}{2}m_f^4 \ln\Biggl(\frac{\mu_f+k_f}{m_f}\Biggr)\Biggr] 
 \protect\label{eos_e_uqm}
\end{eqnarray}

and

\begin{eqnarray}
P_{Q} = -B(\rho)+ \sum_f \frac{1}{4\pi^2} \Biggl[\mu_fk_f\Biggl(\mu_f^2-\frac{5}{2}m_f^2\Biggr)
+ \frac{3}{2}m_f^4 \ln\Biggl(\frac{\mu_f+k_f}{m_f}\Biggr)\Biggr] 
\protect\label{eos_P_uqm}
\end{eqnarray}

where, $m_f$ is the mass of individual quarks and the chemical potential of individual quark is

\begin{eqnarray}
\mu_f=(k_f^2 + m_f^2)^{1/2}
\end{eqnarray} 



The total density is

\begin{eqnarray}
\rho=\sum_f \frac{k_f^3}{3\pi^2}
 \protect\label{density_upq}
\end{eqnarray}

 where, $f$ = u, d and s are the quark flavors. 
 
 The chemical potential equilibrium is given by 

\begin{eqnarray}
\mu_d = \mu_s = \mu_u + \mu_e
\protect\label{Chem_pot_uqm}
\end{eqnarray}

 The individual quark chemical potentials in terms of $\mu_B$ and $\mu_e$ are as follows

\begin{eqnarray}
\mu_u=\frac{1}{3}(\mu_B - 2\mu_e)
\end{eqnarray}

\begin{eqnarray}
\mu_d=\mu_s=\frac{1}{3}(\mu_B + \mu_e)
\end{eqnarray}

The total charge is to be conserved by following the relation

\begin{eqnarray}
\rho_c = \sum_{i} q_i\rho_i =0
\protect\label{charge_neutrality_uqm}
\end{eqnarray}

where, $i$ = u, d, s and e. $q_i$ and $\rho_i$ are individual charge and density of the particles, respectively.

 In the present work, we assume that the surface tension at hadron-quark boundary is sufficiently large and thus phase transition is achieved using Maxwell construction \cite{Maruyama}. Following Maxwell construction, transition from hadronic phase to quark phase occurs with sharp jumps in density when the the pressure and baryon chemical potential of the individual charge neutral phases are equal \cite{Schramm,Lenzi,Bhattacharya,Gomes2019,Han,Ferreira2020, Sen,Sen2} i.e.,

\begin{eqnarray}
\mu_B^H=\mu_B^{Q} 
\end{eqnarray}

and

\begin{eqnarray}
P_H=P_{Q} 
\end{eqnarray}

  We compute the hybrid EoS for different values of the bag pressure by varying $B_{as}$. 
  
   For the outer crust region, we adopted the Baym-Pethick-Sutherland (BPS) EoS \cite{Baym71} and for the inner crust, we have considered the EoS including the pasta phases \cite{Grill14}. 
   Consequently, we proceed to study the structural properties of the HSs in static conditions with the obtained hybrid EoS.

\subsection{Structural Properties of Hybrid Stars}
\label{structure}  
 
 With the obtained hybrid EoS, the structural properties like the gravitational mass ($M$) and the radius ($R$) of the HSs in static conditions are computed by integrating the following Tolman-Oppenheimer-Volkoff (TOV) equations \cite{tov} based on the hydrostatic equilibrium between gravity and the internal pressure of the star.

\begin{eqnarray}
\frac{dP}{dr}=-\frac{G}{r}\frac{\left(\varepsilon+P\right)
\left(M+4\pi r^3 P\right)}{(r-2 GM)},
\label{tov}
\end{eqnarray}

\begin{eqnarray}
\frac{dM}{dr}= 4\pi r^2 \varepsilon,
\label{tov2}
\end{eqnarray} 
 
 The dimensionless tidal deformability ($\Lambda$) is obtained in terms of the mass, radius and the tidal love number ($k_2$) following \cite{Hinderer}. From the deformation of the metric $h_{\alpha \beta}$ in Regge-Wheeler gauge,
 
\begin{eqnarray} 
h_{\alpha\beta}=diag\left[e^{-\nu(r)}H_0,e^{\lambda(r)}H_2,r^2K(r),r^2\sin^2\theta K(r)\right]
Y_{2m}(\theta,\phi)
\label{h}
\end{eqnarray}
 
 the tidal Love number $k_2$ is obtained which in turn gives the tidal deformability parameter $\lambda$ as

\begin{eqnarray} 
\lambda=\frac{2}{2} k_2 R^4
\label{lam}
\end{eqnarray}

 The dimensionless tidal deformability $\Lambda$ is then calculated as a function of Love number, gravitational mass and radius \cite{Hinderer} as.
 
\begin{eqnarray} 
 \Lambda=\frac{2}{3} k_2 (R/M)^5
 \label{Lam}
\end{eqnarray}


\section{Results}
\label{Results}
 
 The EoS for the pure hadronic matter with the six chosen models are calculated individually. For the quark phase, we choose the asymptotic value $B_{as}$ of the bag pressure as 30, 50 and 70 MeV fm$^{-3}$. With these chosen values of $B_{as}$, we obtain the EoS of quark phase using eqs. \ref{eos_e_uqm} and \ref{eos_P_uqm}. With each hadronic EoS we compute the hybrid EoS for each value of $B_{as}$ following Maxwell construction. 

 We compare the pressure as a function of the baryon chemical potential of the both the phases in order to obtain the hadron-quark transition or the crossover points for different values of  $B_{as}$ with each hadronic EoS. The transition chemical potential ($\mu_t$) and the transition pressure ($P_t$) decide the hadron-quark crossover points. We tabulate in table \ref{table_trans} the transition densities of the hadronic ($\rho_t^H$) and quark ($\rho_t^Q$) phases corresponding to the crossover points ($\mu_t$,$P_t$) for different hadronic models and $B_{as}$.
 
\begin{table*}[!ht]
\begin{center}
\caption{Hadron-quark transition densities for different hadronic models and the chosen values of $B_{as}$.}
\setlength{\tabcolsep}{10.0pt}
\begin{center}
\begin{tabular}{cccccccc}
\hline
\hline
\multicolumn{1}{c}{Hadronic Model} &
\multicolumn{1}{c}{$B_{as}$} &
\multicolumn{1}{c}{$\rho_t^H/\rho_0$} &
\multicolumn{1}{c}{$\rho_t^Q/\rho_0$} \\
\multicolumn{1}{c}{} &
\multicolumn{1}{c}{(MeV fm$^{-3}$)} &
\multicolumn{1}{c}{} &
\multicolumn{1}{c}{} \\
\hline
TM1   &30  &1.59  &3.24  \\
      &50  &1.70  &3.40  \\
      &70  &1.83  &3.59  \\
\hline
BSR2  &30  &1.47  &3.24  \\
      &50  &1.56  &3.39  \\
      &70  &1.64  &3.57  \\
\hline
BSR6  &30  &1.48  &3.23  \\
      &50  &1.61  &3.39  \\
      &70  &1.69  &3.57  \\
\hline
GM1   &30  &1.51  &3.23  \\
      &50  &1.64  &3.38  \\
      &70  &1.76  &3.56  \\
\hline
NL3$\omega\rho$4  &30  &1.37  &3.22  \\
                  &50  &1.50  &3.37  \\
                  &70  &1.64  &3.54  \\
\hline
NL3   &30  &1.31  &3.22  \\
      &50  &1.41  &3.36  \\
      &70  &1.51  &3.53  \\
\hline
\hline
\end{tabular}
\end{center}
\protect\label{table_trans}
\end{center}
\end{table*}

 For any hadronic model, the crossover shifts to higher values of chemical potential (transition densities) with higher value of $B_{as}$. For any particular value of $B_{as}$, phase transition is earliest with NL3 model and most delayed in case of TM1 model in terms of transition density. We then proceed to compute the hybrid EoS with the six chosen hadronic EoS for each value of $B_{as}$. The difference in the values of $\rho_t^H$ ($\varepsilon_t^H$) and $\rho_t^Q$ ($\varepsilon_t^Q$) decides the region of phase transition with jump in density according to Maxwell construction.

\begin{figure}[!ht]
\centering
\subfloat[]{\includegraphics[width=0.33\textwidth]{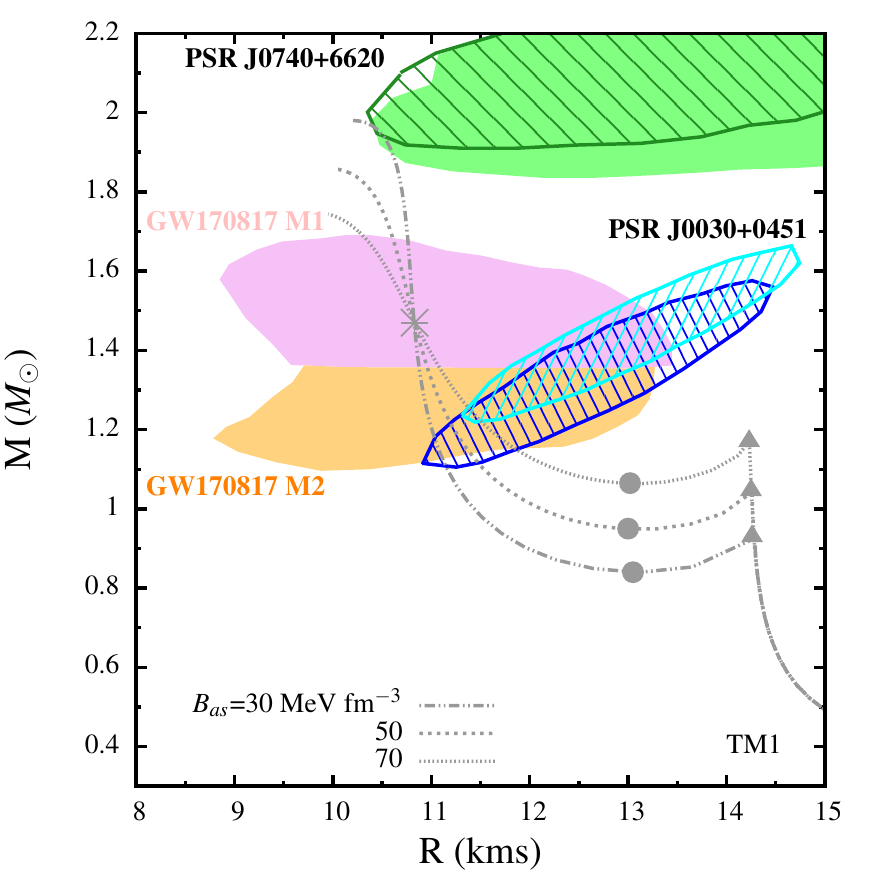}\protect\label{mr_tm1}}
\hfill
\subfloat[]{\includegraphics[width=0.33\textwidth]{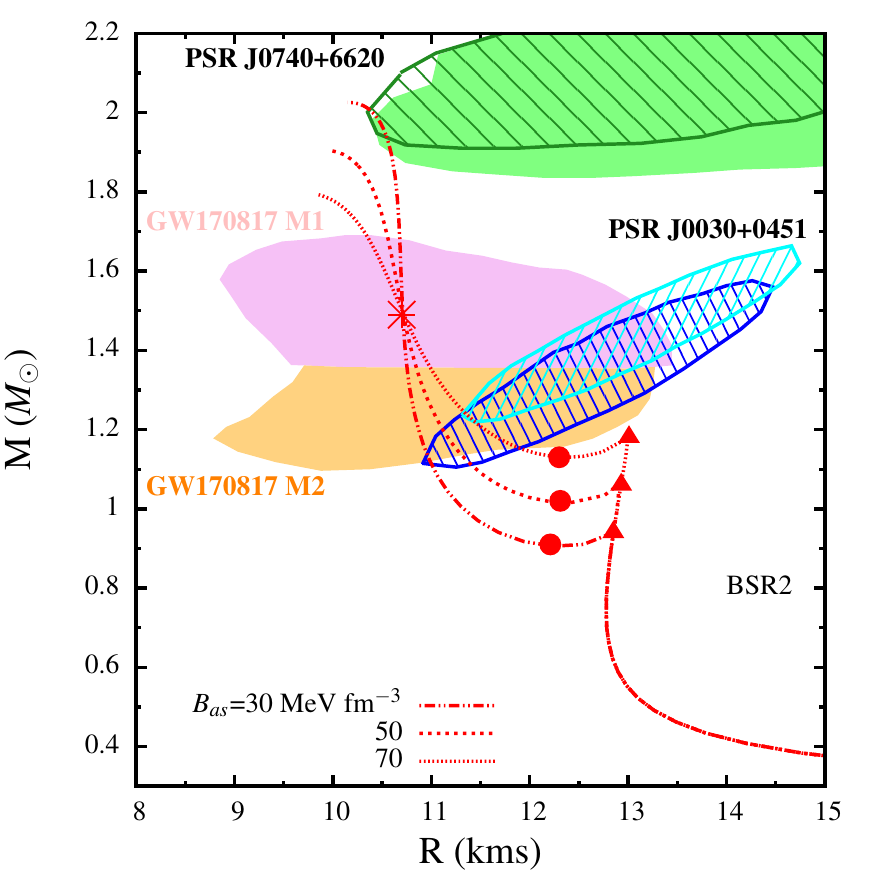}\protect\label{mr_bsr2}}
\hfill
\subfloat[]{\includegraphics[width=0.33\textwidth]{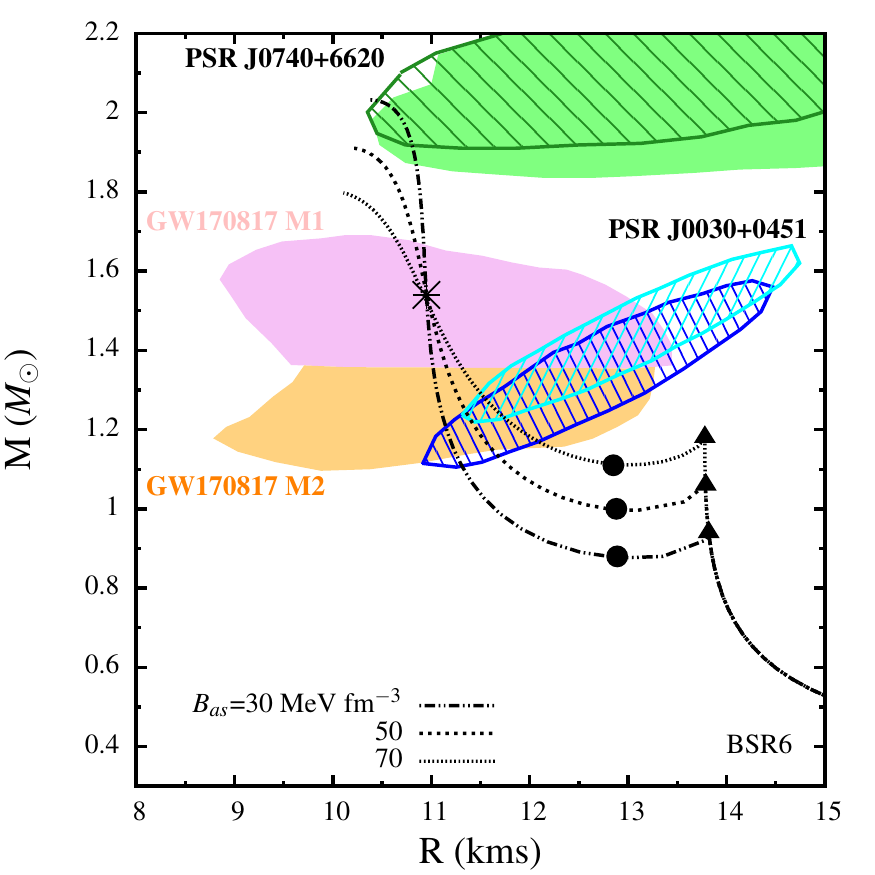}\protect\label{mr_bsr6}}
\caption{\it Mass-radius relationship of static hybrid star with hadronic models (a)TM1, (b)BSR2 and (c)BSR6 and different values of $B_{as}$. Observational limits imposed from the most massive pulsar PSR J0740+6620 ($M = 2.08 \pm 0.07 M_{\odot}$) \cite{Fonseca2021} and $R = 13.7^{+2.6}_{-1.5}$ km \cite{Miller2021} or $R = 12.39^{+1.30}_{-0.98}$ km \cite{Riley2021}) are also indicated. The constraints on $M-R$ plane prescribed from GW170817 \cite{GW170817}) and NICER experiment for PSR J0030+0451 \cite{Riley2019,Miller2019} are also compared. The points marked with asterisks indicate the special points, that with triangles indicate the phase transition and the part of the curves between the triangular and solid dot points indicate the region of instability.}
\label{MR1}
\end{figure}

\begin{figure}[!ht]
\centering
\subfloat[]{\includegraphics[width=0.33\textwidth]{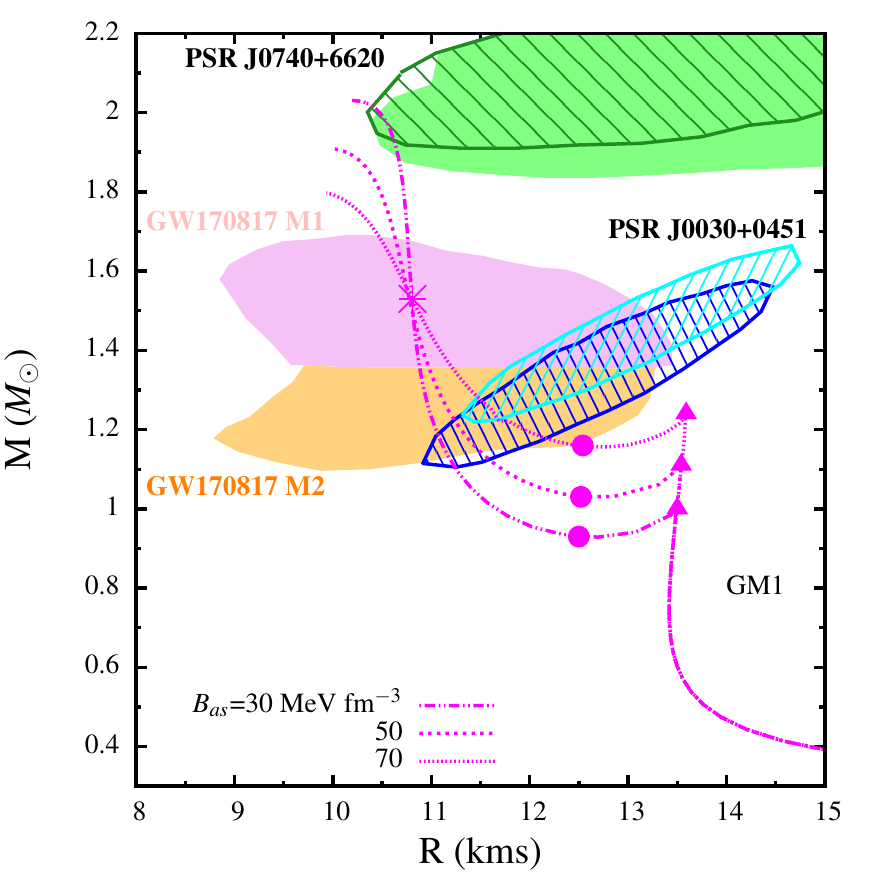}\protect\label{mr_gm1}}
\hfill
\subfloat[]{\includegraphics[width=0.33\textwidth]{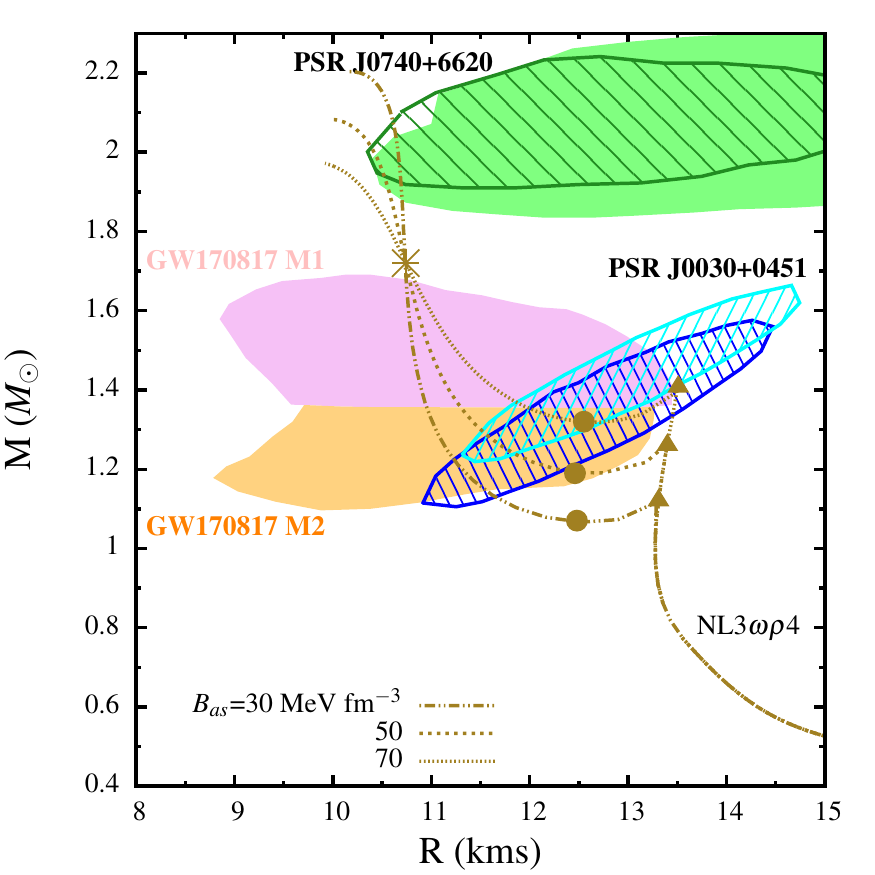}\protect\label{mr_nl3wr4}}
\hfill
\subfloat[]{\includegraphics[width=0.33\textwidth]{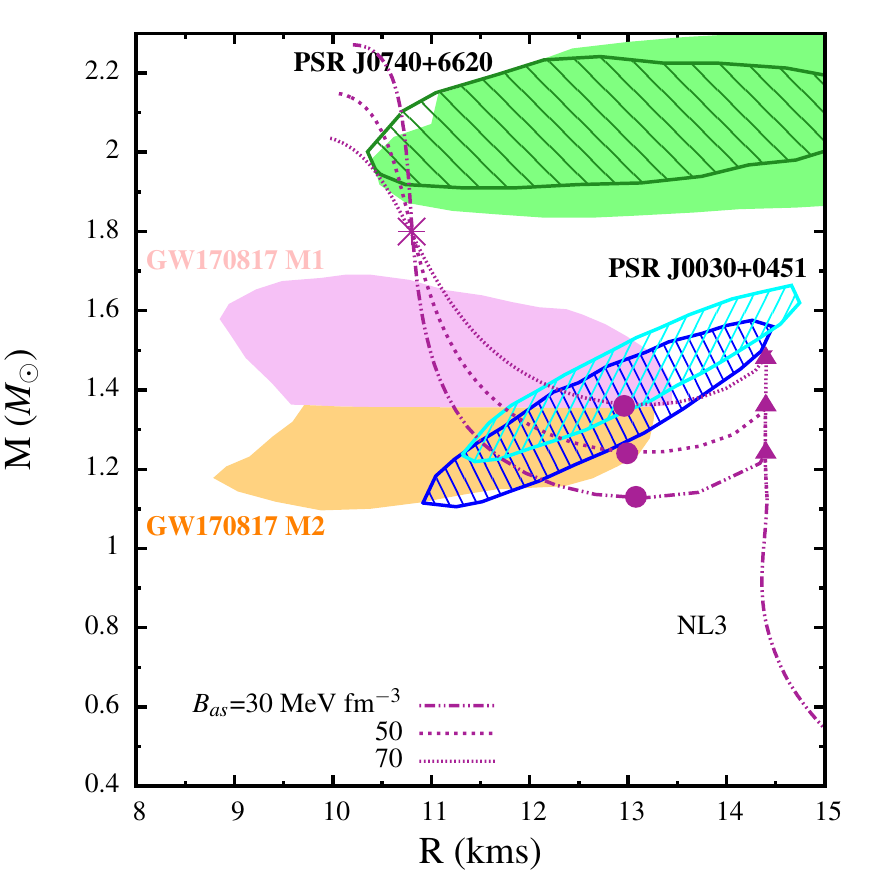}\protect\label{mr_nl3}}
\caption{\it Same as figure \ref{MR1} but with hadronic models (a)GM1, (b)NL3$\omega\rho$4 and (c)NL3.}
\label{MR2}
\end{figure} 
 
 With the obtained hybrid EoS, the structural properties are obtained in static conditions using the TOV equations \cite{tov} as shown in eqs. \ref{tov} and \ref{tov2}. In figures \ref{MR1} and \ref{MR2}, we show the variation of gravitational mass $M$ with radius $R$ of the HSs in static conditions. It is seen that for any particular value of $B_{as}$, the maximum mass ($M_{max}$) of the HSs is highest for NL3 model (2.27 $M_{\odot}$ for $B_{as}$=30 MeV fm$^{-3}$) and lowest for TM1 model (1.98 $M_{\odot}$ for $B_{as}$=30 MeV fm$^{-3}$). For any particular hadronic model, $M_{max}$ of the HSs decreases with increasing values of $B_{as}$. For example, the HSs obtained with GM1 model for the hadronic phase has $M_{max}$=2.03 $M_{\odot}$ for $B_{as}$=30 MeV fm$^{-3}$ and 1.80 $M_{\odot}$ for $B_{as}$=70 MeV fm$^{-3}$. However, the radii of HSs ($R_{max}$) corresponding to $M_{max}$ for different values of $B_{as}$ do not show any substantial change for any particular hadronic model. The mass and radius coordinates ($M_t,R_t$) corresponding to phase transition or the onset of quarks are marked with solid triangles in figures \ref{MR1} and \ref{MR2}. For any particular hadronic model, we find that $M_t$ increases with increasing values of $B_{as}$. This is because the transition is delayed in terms of chemical potential (density) with increasing $B_{as}$, as seen from table \ref{table_trans}. However, all the HS solutions, obtained with different hadronic models show that the value of $M_{max}$ decreases with increasing values of $M_t$ and $B_{as}$. Same relationship between $M_{max}$ and $M_t$ is also reported in other works like \cite{Blaschke2015,Blaschke,Cierniak}.

  The maximum mass \cite{Fonseca2021} and the corresponding radius \cite{Miller2021,Riley2021} constraints from PSR J0740+6620 are found to be satisfied by the HS solutions with hadronic models BSR2, BSR6 and GM1 for $B_{as}$=30 MeV fm$^{-3}$ only. The same $M-R$ constraints from PSR J0740+6620 are satisfied by the HS results with NL3$\omega\rho$4 hadronic model for both $B_{as}$=30 and 50 MeV fm$^{-3}$. Our HS solutions with the NL3 model are consistent with these constraints for all the chosen values of $B_{as}$. Other constraints on the $M-R$ plane from GW170817 \cite{GW170817}) and NICER experiment for PSR J0030+0451 \cite{Riley2019,Miller2019} are not satisfied by the hadronic branches before phase transition. They are, however, well satisfied by all the HS configurations obtained with all the hadronic models for each of the chosen values of $B_{as}$. Overall, our mass-radius results of the HS configurations, as plotted in figures \ref{MR1} and \ref{MR2}, show that the hadronic branch is not massive enough and is incapable of satisfying the NICER data for PSR J0030+0451. This is because phase transition occurs quite early (at low mass) for almost all the considered models (also seen from table \ref{table_trans} in terms of transition density). Therefore the transition mass $M_t$ is quite low while the corresponding radius $R_t$ is large for most of the models considered in this work. Hence the hadronic branch do not reach mass high enough to satisfy the NICER constraint for PSR J0030+0451. This constraint is satisfied with all the stable HS branches or the second stable branch obtained only after phase transition in case of all the models considered in this work. It is also noted from \ref{MR1} and \ref{MR2} that the hadronic branches for none of the model considered in this work can satisfy the GW170817 data either. For all the considered models only the second stable branch after phase transition have successfully satisfied this constraint. The reason is again due to the early phase transition that restricts the hadronic branch to reach the mass and radius corresponding to the GW170817 data.
  
\begin{figure}[!ht]
\centering
\subfloat[]{\includegraphics[width=0.5\textwidth]{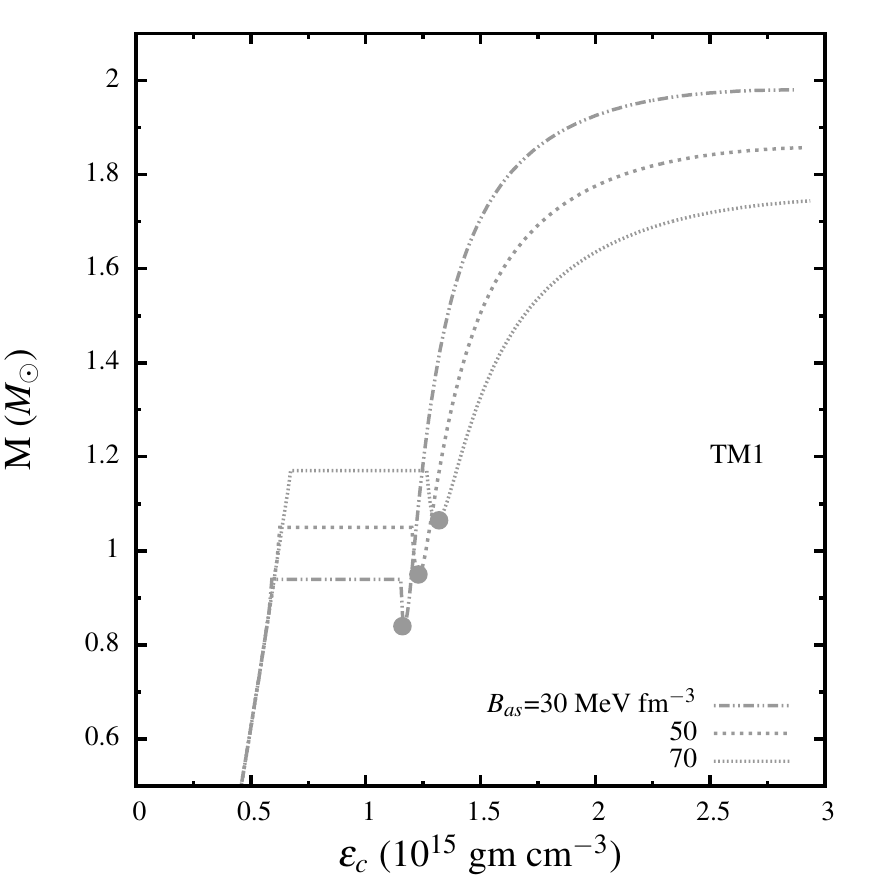}\protect\label{EcM_tm1}}
\hfill
\subfloat[]{\includegraphics[width=0.5\textwidth]{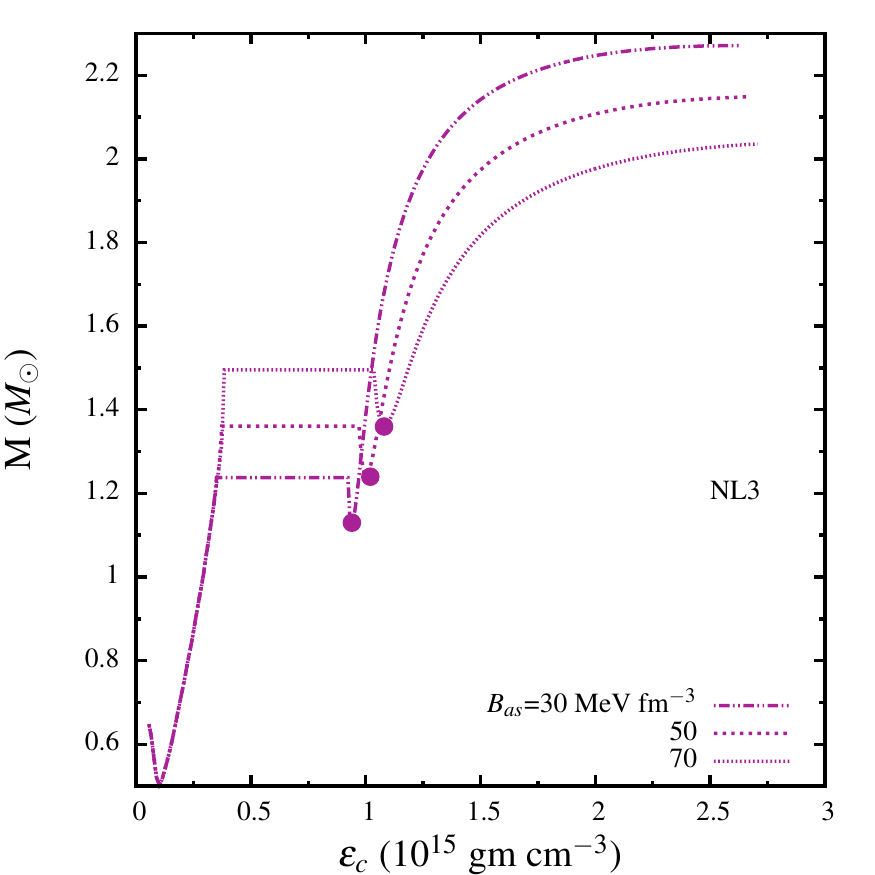}\protect\label{EcM_nl3}}
\caption{Variation of mass with central energy density of hybrid star with hadronic models (a)TM1 and (b)NL3 and different values of $B_{as}$. The points indicate the initiation of stable branch after phase transition.}
\label{EcM}
\end{figure}

 Interestingly, for all the HS configurations presented in figures \ref{MR1} and \ref{MR2}, we obtain twin solutions on two different stable branches separated by an unstable region (between the points marked with solid triangular and circular points in figures \ref{MR1} and \ref{MR2}). This region of instability corresponds to the points when $dM/d\varepsilon_c <0$, where $\varepsilon_c$ is the central energy density. The region is noticed for a considerable density (radius) following phase transition. The solutions become stable again from the circular point onwards which marks the end of the unstable region and the beginning of the second stable branch. This is thus even better reflected if we study the variation of the central energy density $\varepsilon_c$ with respect to mass. In figure \ref{EcM}, we show the same for HSs obtained with hadronic models TM1 and NL3 where we find that the unstable region is clearly seen as a dip in mass with respect to $\varepsilon_c$ following the flat ($M$=constant) region of phase transition.
 
\begin{figure}[!ht]
\centering
\subfloat[]{\includegraphics[width=0.5\textwidth]{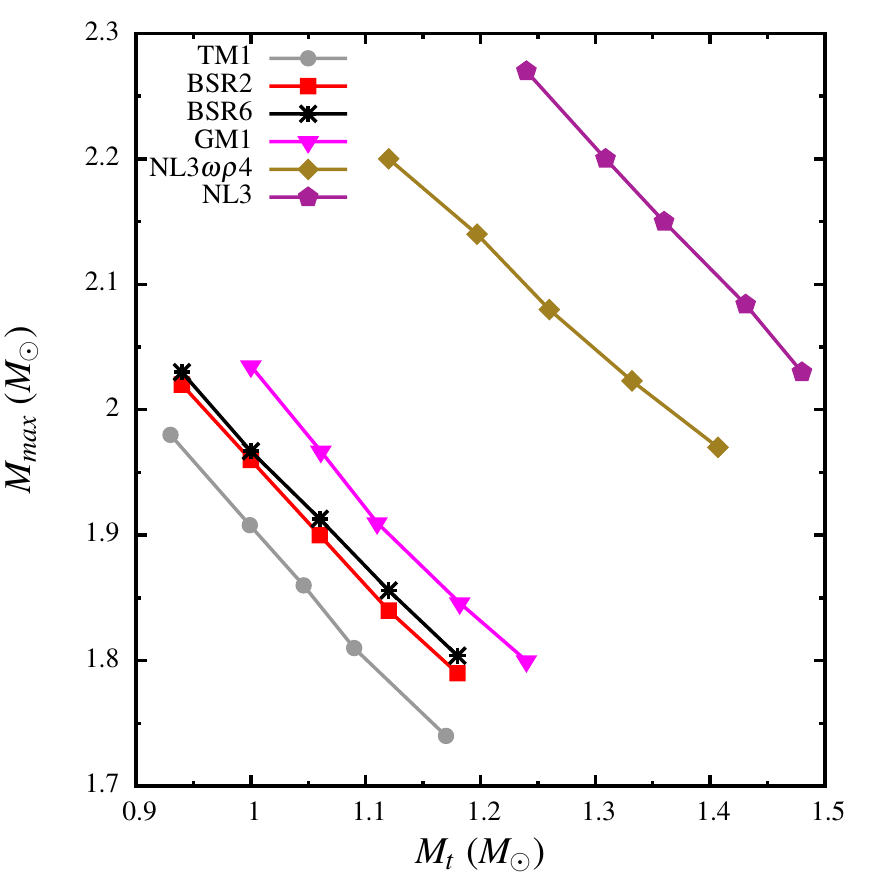}\protect\label{Mmax_Mt}}
\hfill
\subfloat[]{\includegraphics[width=0.5\textwidth]{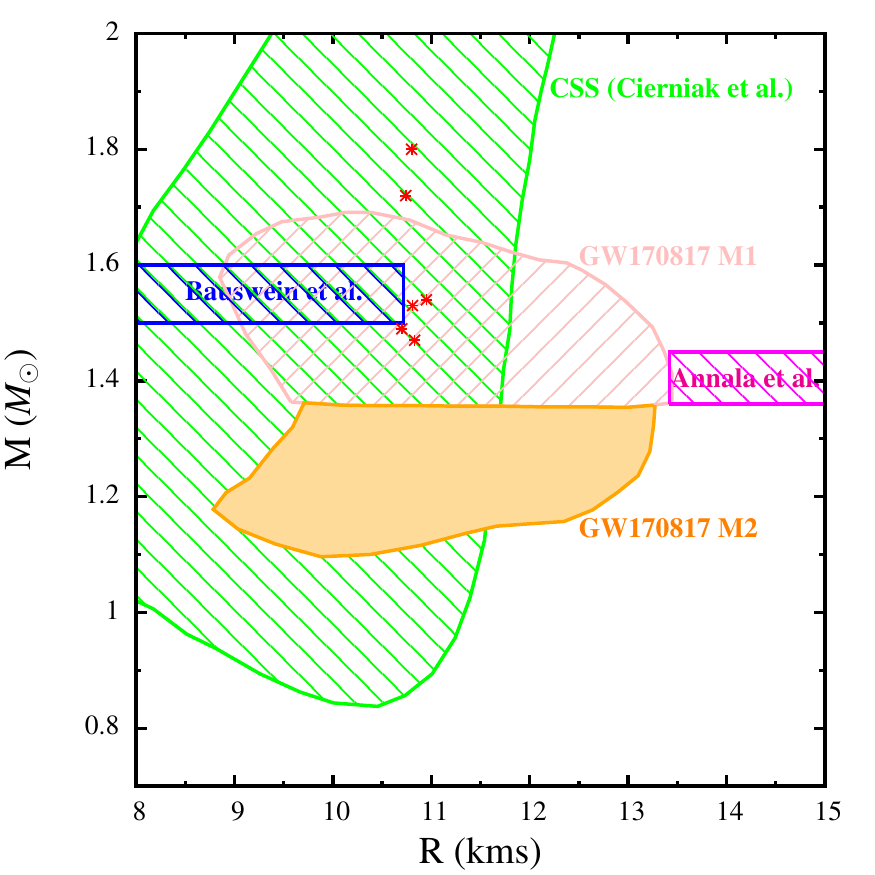}\protect\label{SP}}
\caption{(a)Variation of maximum mass ($M_{max}$) with respect to transition mass ($M_t$) of the hybrid star with different hadronic models. The points represent different the values of $M_{max}$ and $M_t$ for different values of $B_{as}$. (b) Location of special points (asterisks) on the mass-radius plot of hybrid stars with different hadronic models. The possible positions of the special points for the CSS quark model with $C_s^2=$0.7 \cite{Cierniak} is also compared. The allowed \cite{GW170817} and excluded \cite{Bauswein2} regions on the mass-radius plane from GW170817 are also indicated.}
\label{Mmax_Mt_Msp}
\end{figure}

  The decrease in mass is upto a certain point (marked with circular dot) as also in the $M-R$ plots in figures \ref{MR1} and \ref{MR2}. Beyond this point $M$ again increases with $\varepsilon_c$ and also with respect to radius in the $M-R$ plot, forming the second stable branch. The two stable branches are thus characterized by two distinct maxima. Of them the one on the hadronic branch occurs at low mass ($M_t$) and larger radius and corresponds to the point of phase transition (marked by triangles in figures \ref{MR1} and \ref{MR2}). Thus in the hadronic branch the transition point also denote the maxima while in the second stable branch, formed after phase transition, the maxima occurs at high mass ($M_{max}$) and low radius. This maxima denotes the maximum mass of the overall HS. In figure \ref{Mmax_Mt} we show the variation of $M_{max}$ with respect to $M_t$ for the HSs obtained with different hadronic models. In this figure each line is for the HSs obtained with particular hadronic model. The points on any particular line indicate the values of ($M_t$,$M_{max}$) with different values of $B_{as}$. In the $M_t$ vs $M_{max}$ plane we obtain nearly parallel lines for HSs with different hadronic models. The same nature of $M_t$ vs $M_{max}$ is also noticed in other works like \cite{Cierniak}. As also seen in figures \ref{MR1} and \ref{MR2}, we notice in figure \ref{Mmax_Mt} that for HSs with any particular hadronic model, $M_{max}$ decreases with increasing values of $B_{as}$ while the reverse trend is noticed in case of $M_t$ with respect to $B_{as}$.

 Figures \ref{MR1} and \ref{MR2} also show the existence of twin star configurations characterized by points on the $M-R$ plot having same mass but different radii. From \cite{Montana,Christian,Sharifi} it is often seen that the two maxima on the two branches may have close values of mass that exceed 2 $M_{\odot}$ (Category I). In other cases the two maxima differ in mass. In the present work our results mostly belong to the Category III as categorized by \cite{Montana,Christian,Sharifi} in which the maxima on the hadronic branch is just above 1 $M_{\odot}$ while that on the hybrid branch is above 2 $M_{\odot}$ mostly for $B_{as}$=30 MeV fm$^{-3}$. In few cases (figure \ref{MR1}) for the lowest value of $B_{as}$, we have found that our HS configurations belong the Category IV since in such cases the maxima on the hadronic branch is slightly below 1 $M_{\odot}$.

 Another interesting feature of our present study is that the HS configurations obtained with the different values of $B_{as}$ exhibit SPs ($M_{SP},R_{SP}$) on the $M-R$ diagram (indicated by asterisks in figures \ref{MR1} and \ref{MR2}). This phenomenon is noticed for HSs irrespective of the chosen hadronic model. The SPs indicate a small region where the solutions for HSs with different values of $B_{as}$ coincide. Irrespective of the values of $B_{as}$ or $M_t$, the HS solutions intersect at these SPs. This feature of obtaining SP in the $M-R$ diagram of HSs is also noted in works like \cite{Yudin,Cierniak}. In the present work we find that these SPs lie on the second stable branch obtained after phase transition. In \cite{Cierniak} it is seen that SPs can serve as a remarkable tool to interpret the recent multi-messenger observational results as signals for the possible existence of HS branches. Comparing the results of \cite{Cierniak} obtained with the CSS quark model with our results, we find from figure \ref{SP} that our location the SPs with different hadronic models are within the possible region of SPs prescribed by \cite{Cierniak} for constant speed of sound $C_s^2=$0.7 with the CSS quark model. Also our location of the SPs do not violate the excluded regions of the mass-radius plane as prescribed from GW170817 analysis \cite{Bauswein2}. However, our location of SPs do not reach high values of $M_{SP}$ beyond 2 $M_{\odot}$. Therefore the maximum mass constraint region is not satisfied by the $M_{SP}$ values obtained with any of the models in this work. This constraint is, however, satisfied by the maximum mass of our HS configurations mainly with the chosen lowest value of $B_{as}$.
 
\begin{figure}[!ht]
\centering
{\includegraphics[width=0.7\textwidth]{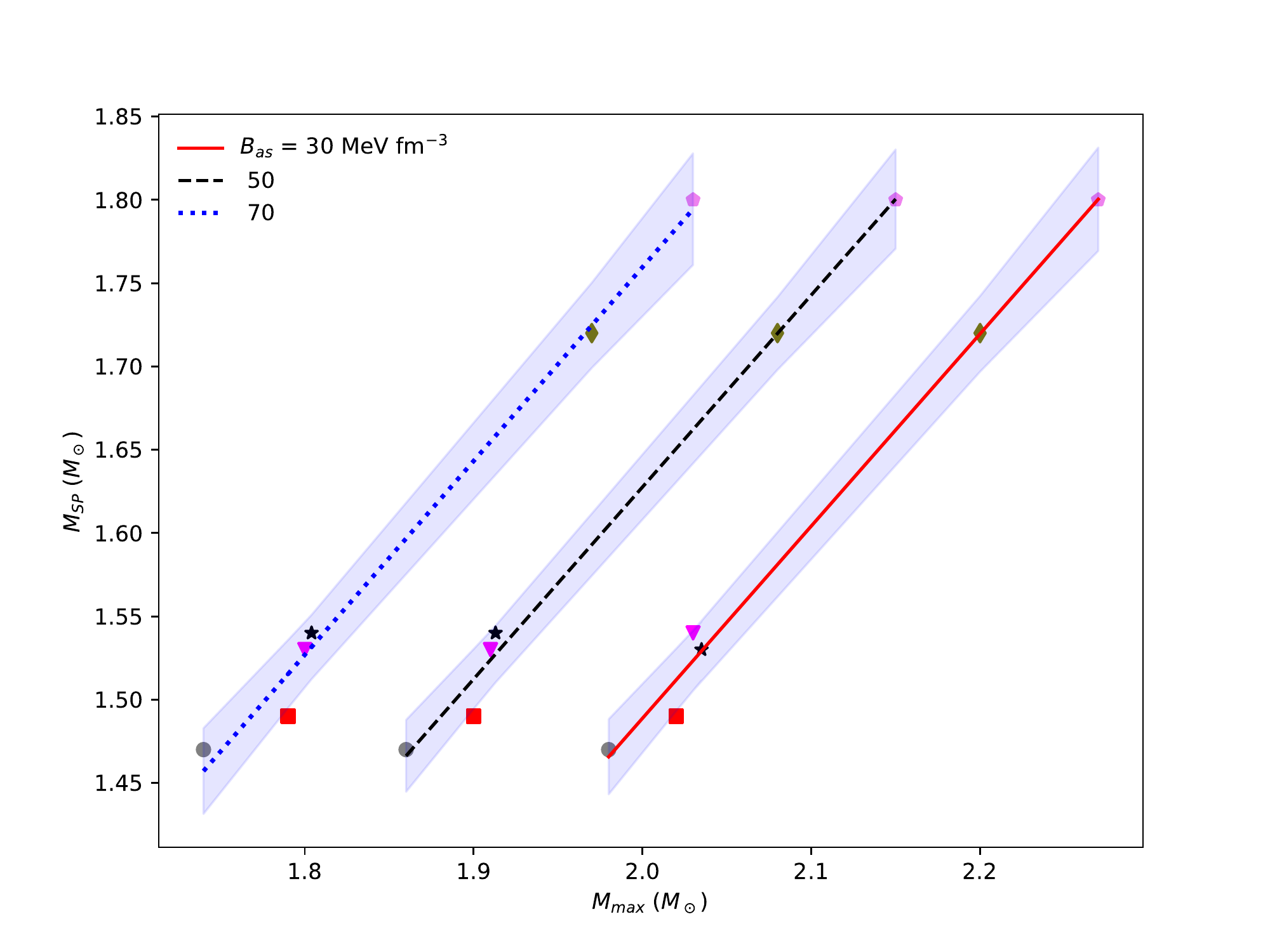}
\caption{Variation of mass corresponding to special point ($M_{SP}$) with respect to maximum mass ($M_{max}$) of the hybrid star with different hadronic models. The different lines represent the fitted functions for different values of $B_{as}$.}
\protect\label{Mmax_Msp}}
\end{figure}

 In the present work the coordinates ($M_{SP},R_{SP}$) vary widely for the HSs with different hadronic models. In figure \ref{Mmax_Msp} we show the variation of the mass corresponding to the SPs ($M_{SP}$) with respect to $M_{max}$ for HSs with different hadronic models for particular values of $B_{as}$.
 
 For any particular value of $B_{as}$, $M_{SP}$ is found to follow a nearly linear (fitted) relationship with $M_{max}$. The particular relation between $M_{SP}$ and $M_{max}$ for different $B_{as}$ is obtained as
  
\begin{eqnarray}
  M_{SP}=-0.81914 + 1.154~M_{max}\rm{,~ for}~B_{as}=30 \rm{~MeV~fm}^{-3}
\label{fit30}              
\end{eqnarray}

\begin{eqnarray}
 M_{SP}=-0.67738 + 1.1525~M_{max}\rm{,~ for}~B_{as}=50 \rm{~MeV~fm}^{-3}
\label{fit50}                            
\end{eqnarray}

and

\begin{eqnarray}
 M_{SP}=-0.56679 + 1.1632~M_{max}\rm{,~ for}~B_{as}=70 \rm{~MeV~fm}^{-3}
\label{fit70}              
\end{eqnarray}

 Thus following eqs. \ref{fit30}, \ref{fit50} and \ref{fit70} we have obtained the constraints on the $M_{max}-M_{SP}$ plane for individual values of $B_{as}$. Such fitted linear relations, given by eqs. \ref{fit30}, \ref{fit50} and \ref{fit70}, can be treated as universal relations in the context of existence of SPs in the case of HSs. These relations do not depend on the transition densities and are found to be satisfied by HS configurations obtained with any of the chosen hadronic models. The lines in figure \ref{Mmax_Msp} are almost parallel to each other and hence the slope is also independent of $B_{as}$. The shaded regions in figure \ref{Mmax_Msp} indicate the uncertainty of the fits. These fitted results are thus almost independent of the hadronic EoS. However, a change in the value of $B_{as}$ causes shift in these fits as $B_{as}$ controls the value of $M_{max}$ of HSs while $M_{SP}$ is independent of $B_{as}$. However, the linearity in the $M_{max}-M_{SP}$ relationship is still seen to be maintained for each value of $B_{as}$. We therefore obtain nearly parallel fitted lines in the $M_{max}-M_{sp}$ plane for different values of $B_{as}$.

\begin{figure}[!ht]
\centering
\subfloat[]{\includegraphics[width=0.33\textwidth]{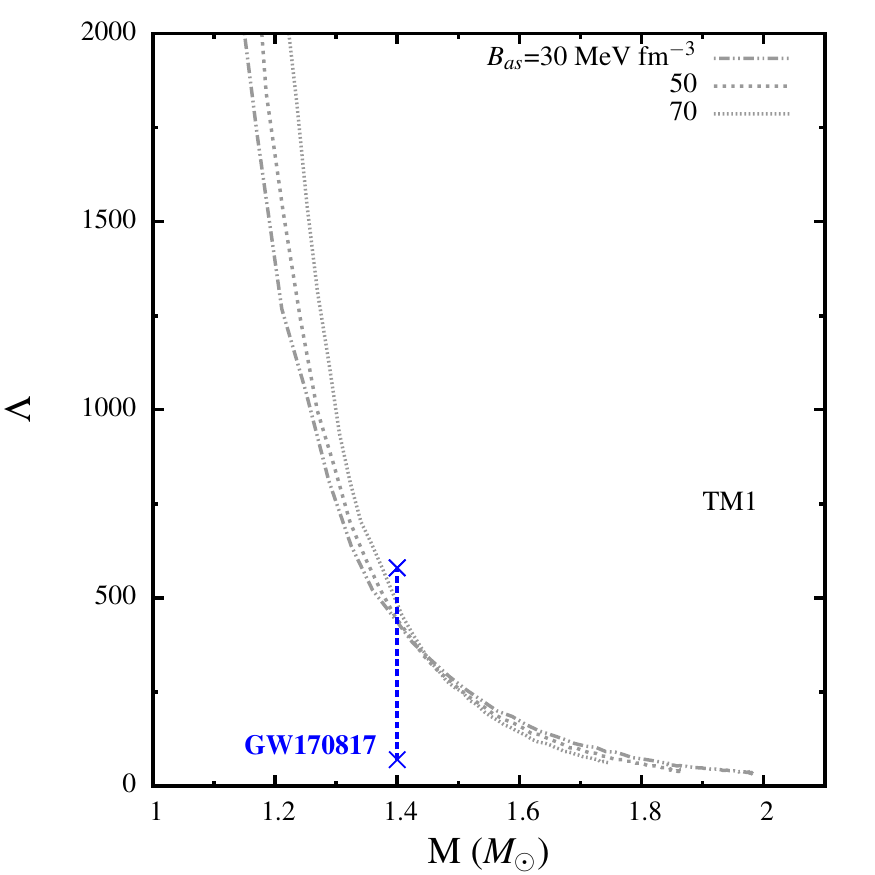}\protect\label{LamM_tm1}}
\hfill
\subfloat[]{\includegraphics[width=0.33\textwidth]{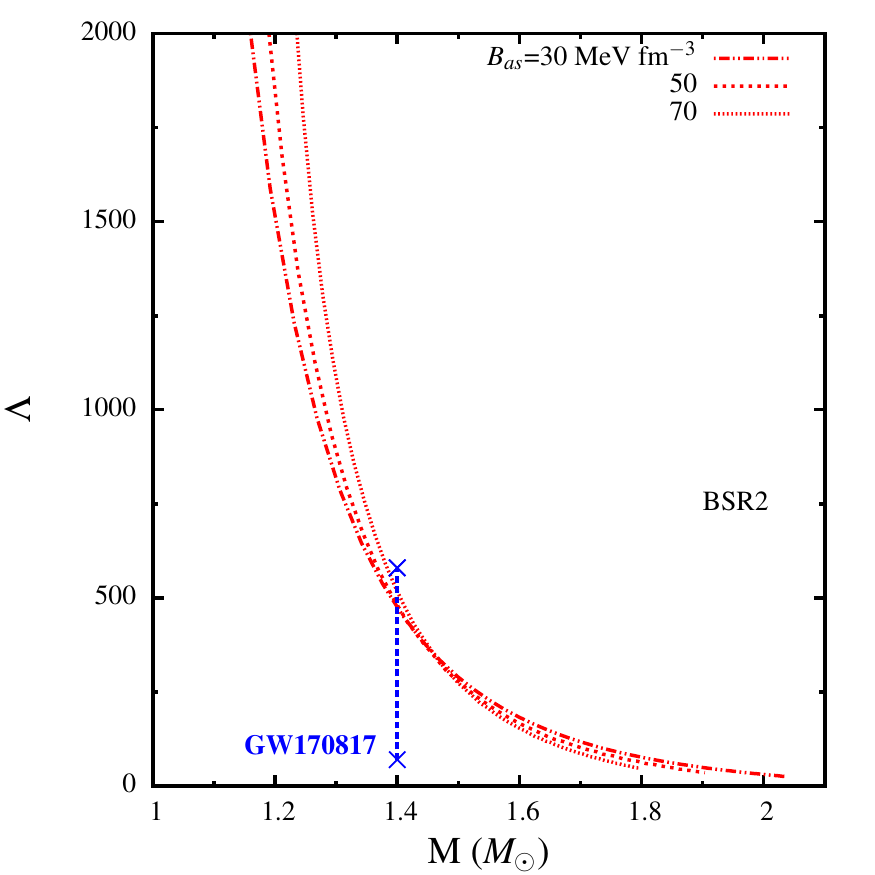}\protect\label{LamM_bsr2}}
\hfill
\subfloat[]{\includegraphics[width=0.33\textwidth]{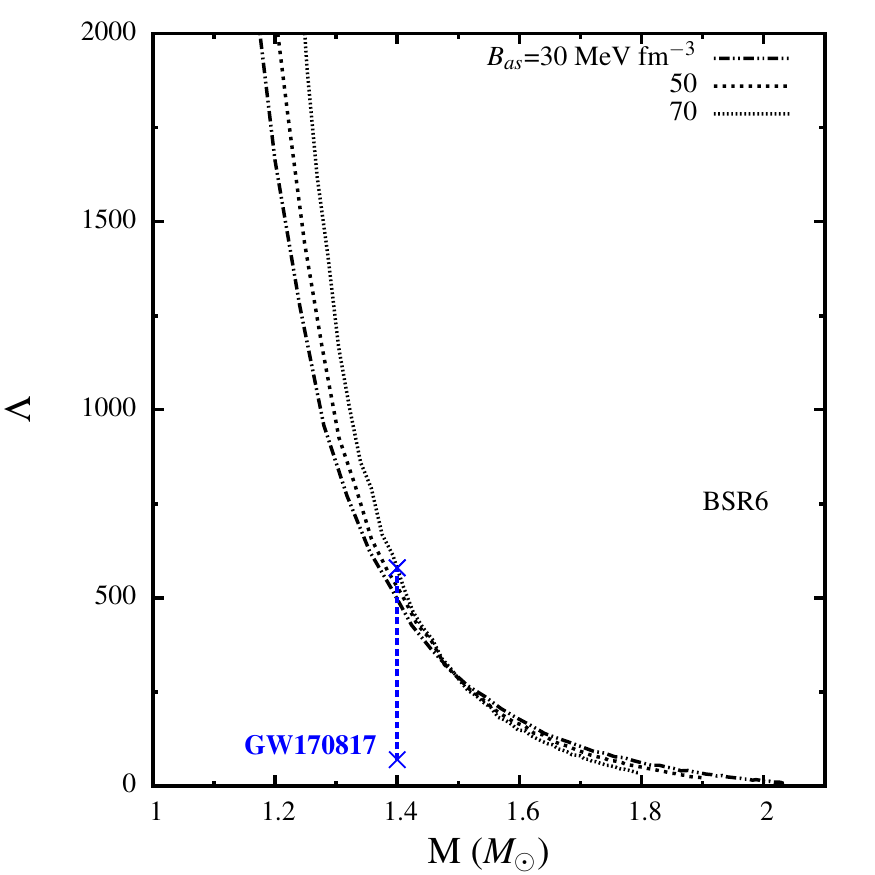}\protect\label{LamM_bsr6}}
\caption{\it Variation of tidal deformability with respect to mass of the hybrid stars with hadronic models (a)TM1, (b)BSR2 and (c)BSR6. and different values of $B_{as}$. Constraint on $\Lambda_{1.4}$ from GW170817 observations ($\Lambda_{1.4}=70-580$ \cite{GW170817}) is also shown.}
\label{LamM1}
\end{figure}

\begin{figure}[!ht]
\centering
\subfloat[]{\includegraphics[width=0.33\textwidth]{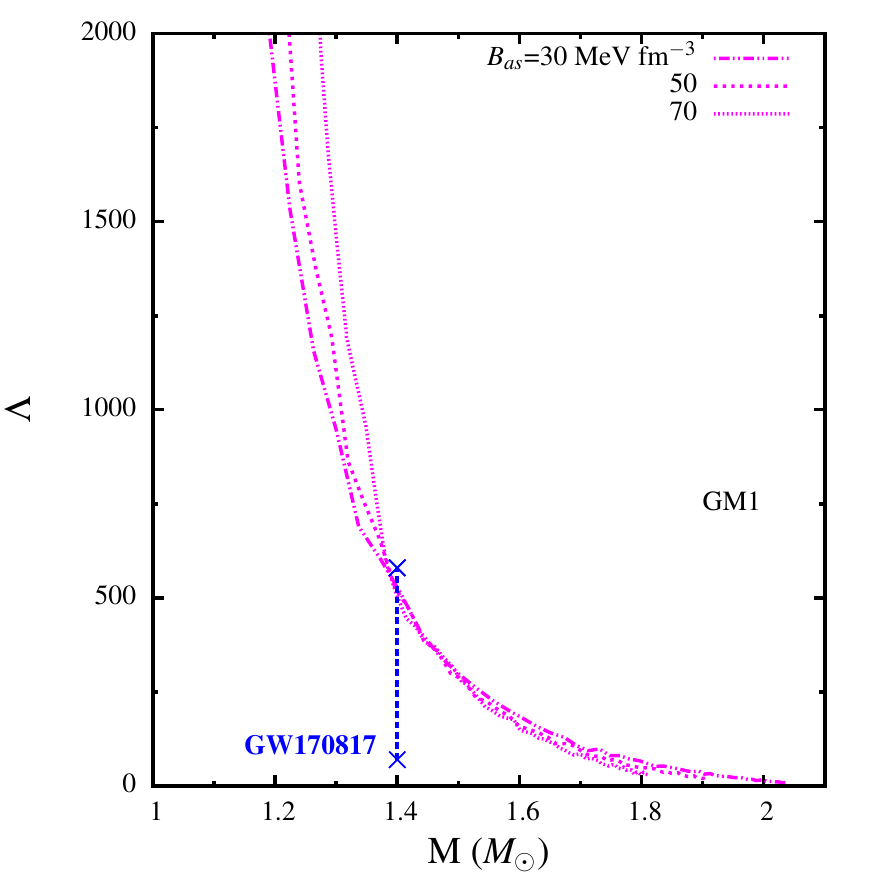}\protect\label{LamM_gm1}}
\hfill
\subfloat[]{\includegraphics[width=0.33\textwidth]{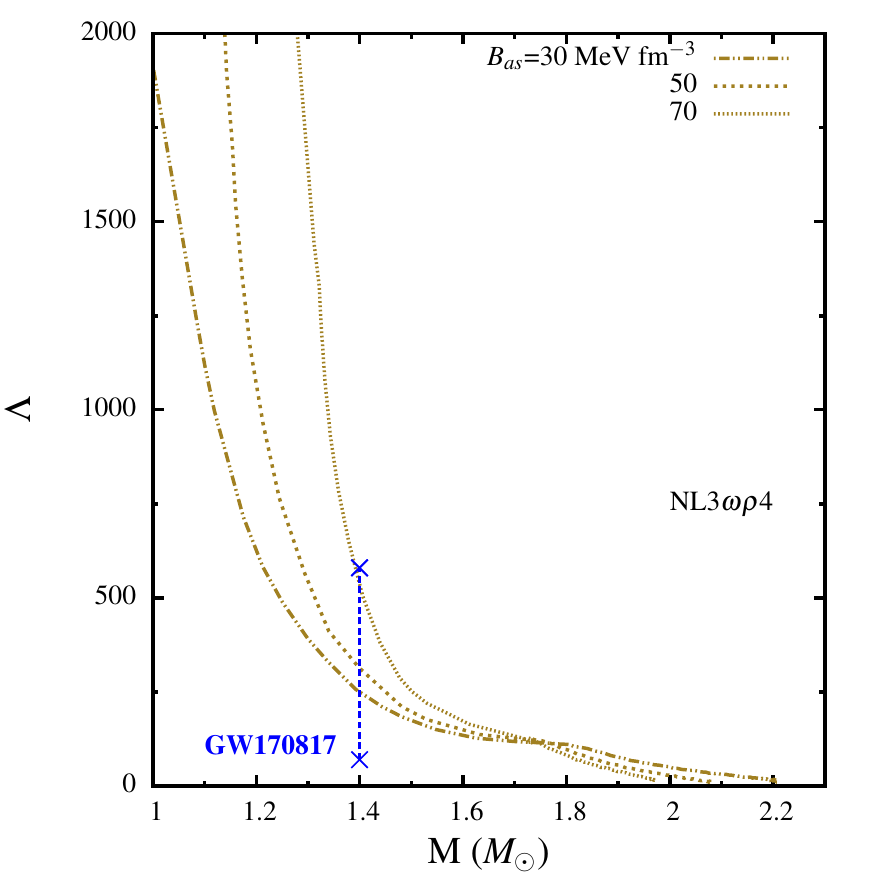}\protect\label{LamM_nl3wr4}}
\hfill
\subfloat[]{\includegraphics[width=0.33\textwidth]{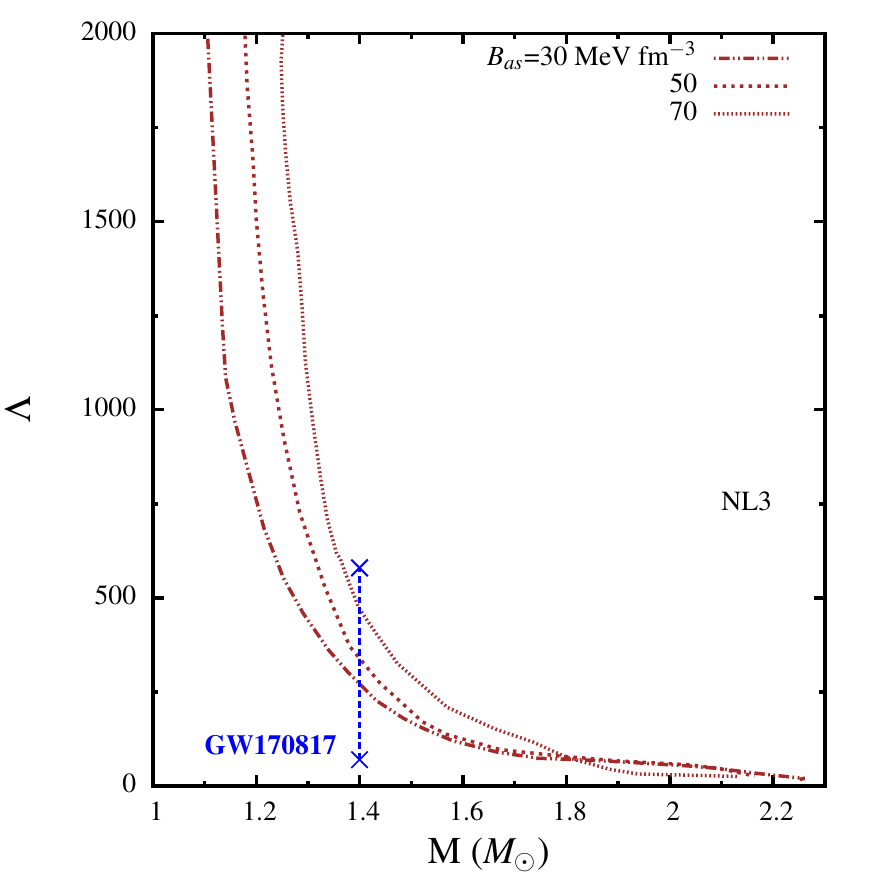}\protect\label{LamM_nl3}}
\caption{\it \it Same as figure \ref{LamM1} but with hadronic models (a)GM1, (b)NL3$\omega\rho$4 and (c)NL3.}
\label{LamM2}
\end{figure}

We next calculate the tidal deformability ($\Lambda$) of the HSs obtained with various hadronic models and the chosen values of $B_{as}$ using eq. \ref{Lam}. In figures \ref{LamM1} and \ref{LamM2}, we show the variation of tidal deformability $\Lambda$ with respect to gravitational mass $M$ of the HSs.

 As expected, $\Lambda$ decreases with $M$ showing that massive stars are less deformed. All the HS configurations obtained with different hadronic models and the chosen values of $B_{as}$ are seen to satisfy the constraint on $\Lambda_{1.4}$ obtained from the data analysis of GW170817 observation \cite{GW170817}. Interestingly, in these plots the $\Lambda-M$ curves for different values of $B_{as}$ either overlap over a region in the vicinity of $M_{SP}$ or converge at $M_{SP}$. This feature is similar to that obtained in the $M-R$ plots in figures \ref{MR1} and \ref{MR2}.


\section{Summary and Conclusion}
\label{Conclusion}

 The possibility of hadron-quark phase transition in NS cores and the formation of HSs are investigated in the present work. For the purpose six different RMF hadronic models are employed for the hadronic phase while the density dependent MIT Bag model is adopted for the pure quark phase. The density dependence of the bag pressure is considered for different values of $B_{as}$. We studied the structural properties of the HSs in the light of the various astrophysical constraints on them. The results highlight the possible formation of twin stars with special emphasis on the existence of SPs on the mass-radius diagram of the HSs with different values of $B_{as}$ using each of the six hadronic models. For each value of $B_{as}$, we obtain nearly hadronic model independent relations between $M_{SP}$ and $M_{max}$ in almost linear forms. Thus these relations can be considered to be universal relations in the context of SPs on the mass-radius relationship of HSs. The calculated HS properties satisfy the present day astrophysical constraints on the $M-R$ relation obtained from PSR J0740+6620, GW170817, and NICER experiment for PSR J0030+0451 and also that on $\Lambda_{1.4}$ obtained from GW170817 data analysis.


\section*{Acknowledgement}

N. Alam and G. Chaudhuri acknowledge the support of ``IFCPAR/CEFIPRA" project 5804-3.

\end{document}